\documentclass[journal]{IEEEtran}

\usepackage{cite}
\usepackage{graphicx}
\usepackage[cmex10]{amsmath}
\usepackage{amssymb}
\usepackage{booktabs}
\usepackage{threeparttable}
\usepackage{algorithm}
\usepackage{algpseudocode}
\usepackage{amsfonts}
\usepackage{xcolor}
\usepackage{color}

\usepackage{tabularx} %

\usepackage{enumitem}
\usepackage{framed}

\usepackage{dblfloatfix} %

\usepackage{soul}

\usepackage{cancel} %

\usepackage{verbatim}%

\usepackage{makecell}

\usepackage{empheq} %

\usepackage{underoverlap} %

\usepackage{hyperref}
\hypersetup{pdftex,colorlinks=true,allcolors=black}

\usepackage{calc}

\usepackage{textcomp}
\def\BibTeX{{\rm B\kern-.05em{\sc i\kern-.025em b}\kern-.08em
		T\kern-.1667em\lower.7ex\hbox{E}\kern-.125emX}}

\newcommand{\HFs}{hopping frequencies}

\newcommand{\mj}{\mathsf{j}}

\newcommand{\myRound}[1]{\lfloor#1\rceil}

\newcommand{\myFloor}[1]{\lfloor#1\rfloor}

\begin{document}

\title{Frequency-Hopping MIMO Radar-Based Communications: An Overview}

\author{{
	Kai Wu,
	J. Andrew Zhang,~\IEEEmembership{Senior Member,~IEEE}, %
	Xiaojing Huang,~\IEEEmembership{Senior Member,~IEEE}, and\\}
	Y. Jay Guo,~\IEEEmembership{Fellow,~IEEE}
	\thanks{Manuscript received XXX, XX, 2020.
		This work is partially funded by the NSW Defence Innovation Network and the NSW State Government through Pilot Project grant DINPP-19-20 10.01}
	\thanks{K. Wu, J. A. Zhang, X. Huang and Y. J. Guo are with the Global Big Data Technologies Centre, University of Technology Sydney, Sydney, NSW 2007, Australia (e-mail: kai.wu@uts.edu.au; andrew.zhang@uts.edu.au; xiaojing.huang@uts.edu.au;   jay.guo@uts.edu.au).}%
}

\maketitle

\begin{abstract}
	
	Enabled by the advancement in radio frequency technologies, the convergence of radar and communication systems becomes increasingly promising and is envisioned as a key feature of future 6G networks. Recently, the frequency-hopping (FH) MIMO radar is introduced to underlay dual-function radar-communication (DFRC) systems. Superior to many previous radar-centric DFRC designs, the symbol rate of FH-MIMO radar-based DFRC (FH-MIMO DFRC) can exceed the radar pulse repetition frequency. 
	{However, many practical issues, particularly those regarding effective data communications, are unexplored/unsolved.}
	To promote the awareness and general understanding of the novel DFRC, this article is devoted to providing a timely introduction of FH-MIMO DFRC. We comprehensively review many essential aspects of the novel DFRC: channel/signal models, signaling strategies, modulation/demodulation processing and channel estimation methods, to name a few. 
	We also highlight major remaining issues in FH-MIMO DFRC and suggest potential solutions to shed light on future research works.
	 
\end{abstract}


\section{Introduction}\label{sec: introduction}
Enabled by the advancement in radio frequency (RF) technologies and signal processing, the convergence of multi-functional RF systems becomes increasingly promising and is envisioned as a key feature of future 6G networks~\cite{6G_visionNewParadigm}. 
Among numerous RF systems, including wireless communications, radio sensing, mobile computing, localization, etc., the former two have achieved significant progress recently on their integration. This is evidenced by several timely overview, survey and tutorial papers~\cite{DFRC_CoexistenceOverview2019SigMagLeZheng,Kai_rahman2020enablingSurvey,FanLiu2019Overview,DFRC_SP_Mag2019Amin_Aboutanios}.

Driven by the spectrum scarcity cost saving, co-existence between the two RF functions has been investigated in the past few years, with focus on mitigating the interference between the two functions~\cite{DFRC_CoexistenceOverview2019SigMagLeZheng}. Thanks to the shared commonalities in terms of signal processing algorithms,
hardware and, to some extent, system architecture, joint communication and radar/radio sensing (JCAS), also referred to as dual-function radar-communication (DFRC), is emerging as an effective solution for integrating wireless communication and radio sensing~\cite{Kai_rahman2020enablingSurvey}. Substantially different from the co-existence of the two RF functions, JCAS/DFRC aims to design and use a single transmitted signal for both communication and
sensing, enabling a majority of the transmitter modules and receiver hardware to be shared.

The design of JCAS/DFRC can be communication-centric or radar-centric \cite{FanLiu2019Overview}.
The former performs radar sensing using ubiquitous communication signals, e.g., IEEE 802.11p \cite{surender2011uwb} and IEEE 802.11ad \cite{JointRC_AdaptiveWaveform2019TSP_Robert}, whereas the later embeds information bits into existing radar waveform or specifically optimized dual-function waveforms \cite{FanLiu2019Overview}.
Regarding the communication-centric JCAS, it is worth mentioning a recently proposed perceptive mobile network (PMN)~\cite{Kai_rahman2020enablingSurvey}. Integrating sensing function into mobile networks, the PMN is envisaged to revolutionize future 5G and beyond networks by offering ubiquitous sensing capabilities for numerous smart applications, such as smart city/factory. Compared with the communication-centric counterpart, radar-centric designs, which has also been typically referred to as DFRC, generally have superior radar sensing performance, given the sensing-dedicated waveform.

Driven by automotive applications, early DFRC tended to employ frequency modulated radar waveforms \cite{barrenechea2007fmcw}.
The advancement of MIMO radars has made the recent DFRC designs leaning towards this new radar paradigm \cite{DFRC_SigStrategyOverview2016AESMag,DFRC_SP_Mag2019Amin_Aboutanios,FanLiu2019Overview}.
Some researchers optimize the beam pattern of a MIMO radar to perform conventional modulations, such as phase shift keying (PSK) and amplitude shift keying, using sidelobes in the MIMO radiation patterns;
others optimize radar waveform to perform non-traditional modulations, such as waveform shuffling 
and code shift keying (see \cite{DFRC_SigStrategyOverview2016AESMag,DFRC_SP_Mag2019Amin_Aboutanios} and the references therein).  
Most previous DFRC works embed one communication symbol within one or multiple radar pulses; hence the communication symbol rate is limited by the pulse repetition frequency (PRF) of the underlying radar.

\begin{figure*}[!t]
	\centerline{\includegraphics[width=130mm]{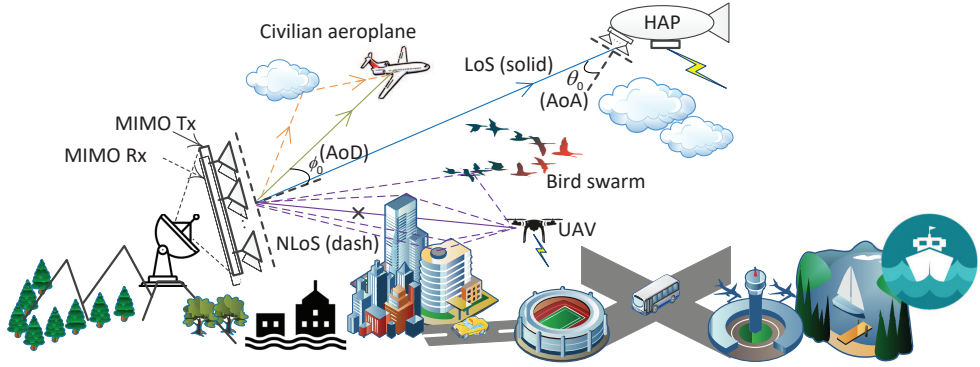}}%
	\caption{A promising scenario of FH-MIMO DFRC, where an FH-MIMO radar is placed in a high-altitude platform to perform air surveillance and meanwhile communicate with an aircraft that can be a UAV, a CAP or an HAP.
	}
	\label{fig: DFRC system}
	\vspace{-10pt}
\end{figure*}

In some recently reported research \cite{DFRC_AmbiguityFunc2018Amin,FH_MIMO_Radar2019_RadarConf,DFRC_DPSK2019Amin,DFRC_cskCPM,DFRC_FHcodeSel2018,DFRC_pskDPSKcpm2020}, frequency-hopping MIMO (FH-MIMO) radar is introduced to DFRC, which breaks the above limit and substantially increases the communication symbol rate to multiples of (e.g., $ 15 $ times) PRF. 
Pioneering in conceiving the novel DFRC architecture based on FH-MIMO radar (FH-MIMO DFRC), the research reported in \cite{DFRC_AmbiguityFunc2018Amin,FH_MIMO_Radar2019_RadarConf,DFRC_DPSK2019Amin,DFRC_cskCPM,DFRC_FHcodeSel2018,DFRC_pskDPSKcpm2020} focuses on analyzing the impact of information modulation on the radar ranging performance. 
Yet, little attention is paid to effective implementation of data communications. {Some latest research reported in \cite{Kai_channelDFRC_ICC2020,Kai_JCASchannelTCOM_ArXiv,Kai_JCASsecureTWC_ArXiv} develops methods to address the practical issues, e.g., channel estimation and synchronization, in FH-MIMO DFRC.}

Motivated by the rapidly increasing interest in DFRC and the lack of awareness and general understanding of FH-MIMO DFRC, this article aims to provide a timely introduction of the novel DFRC architecture.
We start by discussing the potential applications and channel scenarios of FH-MIMO DFRC and also the commonly used signal model in the literature. 
Then, we survey the existing signaling strategies for FH-MIMO DFRC, illustrate their modulation and demodulation methods, and also compare the signaling strategies from various aspects. {We further discuss the issue of channel estimation in FH-MIMO DFRC, reporting recently developed solutions.} 
Finally, we highlight the major issues unsolved in FH-MIMO DFRC and suggest potential solutions to shed light on future research directions.   
{It is noteworthy that this article provides a better coverage of the communication aspects, of FH-MIMO DFRC, which are rarely dealt with in prior works.}

\section{Scenarios and Signal Model for FH-MIMO DFRC}\label{sec: signal model}

FH-MIMO radar is pulse-based and hence is likely to be employed in
 applications requiring a large range coverage, such as long-range air-surveillance~\cite{book_radarWaveform2012Gini}. 
Besides, such radar is generally placed in high altitudes, ``looking'' ahead and above, so as to reduce ground clutters and co-frequency interference from terrestrial wireless systems. 
Therefore, FH-MIMO DFRC is promising for providing ground-to-air (G2A) communications for different types of aircraft, as depicted in Fig. \ref{fig: DFRC system}. 
The aircraft can be an unmanned aerial vehicle (UAV), a civilian aeroplane (CAP) or a high-altitude platform (HAP). 
{With careful designs, FH-MIMO DFRC can greatly benefit the underlying RF functions, as detailed below. }

	{\textit{Benefiting Radar:~}FH-MIMO DFRC could potentially provide a radar with secure, long-range, low-latency, and high-speed wireless broadband connections to airplanes or warships over very wide areas (hundreds to thousands of kilometers, at least one-way link).
	The communication links provided by FH-MIMO DFRC are achieved via the careful reuse of the powerful radar infrastructures, without compromising the radar capabilities in detecting, tracking and identifying targets. Compared with typical communication systems, modern/future radar systems have transmitters with much higher power, much larger antenna arrays with very narrow beams and beam steering capability, and receivers with much higher sensitivity. Therefore, FH-MIMO DFRC can provide almost unparalleled communication links that typical communication systems, such as VHF radio and satellite, can hardly provide. 	
	Besides, the integrated solution can achieve significant saving on cost, size, and weight, compared to having separate wireless communication and radar devices.}

\textit{Benefiting Mobile Communications:~}The G2A link provided by FH-MIMO DFRC may also help support seamless wireless coverage, as expected to be realized by future 6G networks. In particular, FH-MIMO DFRC can
contribute to building the integrated space and terrestrial network (ISTN) which is envisioned to be at the core of 6G communication systems \cite{6G_xiaojingPlaneAidedNetw2019TVTmag}. To this end, employing satellites, in combination with FH-MIMO DFRC, may provide a more cost-effective solution to providing wireless connectivity for people and vehicles in remote rural areas and in the air, as well as at sea,
as compared with the solutions solely relying on satellites.
Against the above potential scenarios, we depict the channel models suitable for FH-MIMO DFRC below.

\subsection{Channel Model Suitable for FH-MIMO DFRC} \label{subsec: channel model}

The channel distribution for FH-MIMO DFRC can vary with the altitude of an airborne user equipment (UE). The typical altitudes of UAVs, CAPs and HAPs are $ 10^3 $ m, $ 10\times 10^3 $ m and $ 20\times 10^3 $ m above the sea level, respectively \cite{Channel_UAVchannel_zeng2019accessing}. 
Therefore, in FH-MIMO DFRC, the following flat-fading channel distributions are likely to be present.   

\begin{enumerate}
	\item Rayleigh channel is likely to exist between the radar and a UAV, particularly when the line-of-sight (LoS) path is blocked by a high-rise building and there exist numerous scattering paths; see Fig. \ref{fig: DFRC system}.

	\item Rician channel can be common between the radar and a CAP, where only few NLoS paths are present and are possibly much weaker than the LoS path;
	
	\item Additive white Gaussian noise (AWGN) channel can prevail between the radar and an HAP where the LoS dominates the channel with negligible NLoS paths.
		  
\end{enumerate}

The geometric Saleh-Valenzuela (SV) channel model~\cite{SVchannel_beamspaceChannel_LLD}, with flexible parameters, can be adapted for depicting the three distributions.
For ease of illustration, assume that a single antenna is equipped at the UE, which is also widely considered in the related works~\cite{DFRC_AmbiguityFunc2018Amin,FH_MIMO_Radar2019_RadarConf,DFRC_DPSK2019Amin,DFRC_cskCPM,DFRC_FHcodeSel2018,DFRC_pskDPSKcpm2020,Kai_channelDFRC_ICC2020,Kai_JCASchannelTCOM_ArXiv}. 
Accordingly, the SV channel model can be given by
\begin{align} \label{eq: channel model}
	\mathbf{h} = \sum_{p=0}^{P-1} \beta_p \mathbf{a}_{M}(u_p)
\end{align}
where $ \mathbf{a}_M(u_p) $ denotes the $ M $-dimensional steering vector of the radar transmitter array in the direction of $ u_p $. 
The $ m $-th element of $ \mathbf{a}_{M}(u_p) $ is $ e^{-\mj m u_p}~(m=0,1,\cdots,M-1) $.
Here, $ u_p $ is the beamspace-domain angle-of-departure (AoD)~\cite{SVchannel_beamspaceChannel_LLD} which is given by $ \pi\sin\phi_p $ with $ \phi_p $ denoting the physical direction; see Fig. \ref{fig: DFRC system} for the case of $ p=0 $.

	Depending on the values of $ P $ and $ \beta_{p}~\forall p $, the SV channel can be simplified to the three classical channels illustrated above. Specifically, the SV channel model can become 
\begin{enumerate}
	\item Rayleigh channel, given $ P\gg 1 $ and $ \mathrm{var}\{\beta_p\} \approx \mathrm{var}\{\beta_{p'}\}~\forall p,p' $;
	
	\item Rician channel, given $ P>1 $ and $ \mathrm{var}\{\beta_0\} \gg \mathrm{var}\{\beta_{p}\}~\forall p>0 $; and
	
	\item AWGN channel, given $ P=1 $.	
\end{enumerate}
So far, most of works only considered AWGN channels for MIMO-DFRC, with exceptions in, e.g., \cite{Kai_JCASchannelTCOM_ArXiv}.

\subsection{Signal Model}

\begin{figure*}[!t]
	\centerline{\includegraphics[width=110mm]{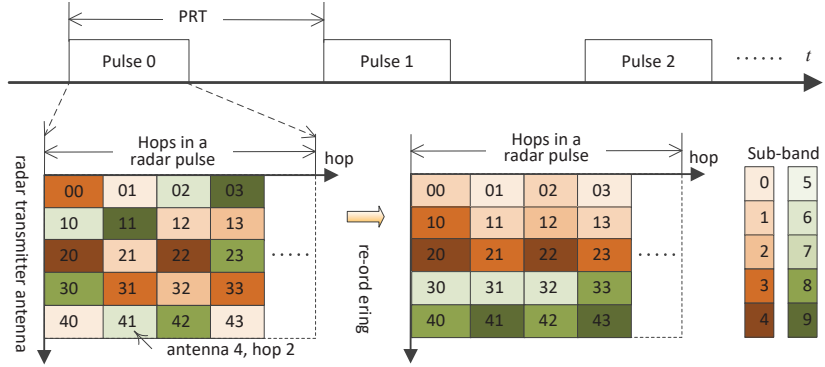}}%
	\caption{Illustration of the signal structure of an FH-MIMO radar, where each radar pulse is divided into multiple hops, each hop randomly selects $ M $ (out of $ K $ in total) sub-bands as hopping frequencies, one for each antenna. The color scale indicates sub-band.}
	\label{fig: signal model}
	\vspace{-10pt}
\end{figure*}

The FH-MIMO radar is based on fast frequency hopping. Namely, each pulse is divided into $ H $ sub-pulses, also referred to as {hops} \cite{DFRC_FHcodeSel2018}. 
The centroid frequency of the transmitted signal changes randomly across hops and antennas, as illustrated in Fig. \ref{fig: signal model}. 
Denote the radar bandwidth as $ B $. By dividing the frequency band evenly into $ K $ sub-bands, each sub-band has the bandwidth of $ B/K $.
Out of the $ K $ centroid frequencies, $ M(<K) $
frequencies are selected to be the \HFs~per hop, one for each antenna. 
Let $ k_{hm} $ denote the index of the sub-band selected by antenna $ m $ at hop $ h $.
To ensure the waveform orthogonality of FH-MIMO radar, the following constraints are imposed for the radar \cite{Amb_FH_MIMO2008TSP},
\begin{align}\label{eq: waveform orthogonlity}
	k_{hm}\ne k_{hm'}~(\forall m\ne m',~\forall h);~\Delta\triangleq{BT}/{K}\in \mathbb{I}_{+},
\end{align}
where $ T $ is the time duration of a hop.  

At hop $ h $, the $ m $-th antenna of the radar transmitter transmits a single-tone signal, i.e.,
\begin{align}\label{eq: transmitted signal p h m}
	{s}_{hm}(i) =	e^{\mj2\pi ik_{hm} \Delta},~i=0,1,\cdots,L-1,
\end{align}
where $ L=\myRound{T/T_{\mathrm{s}}} $ is the number of samples per hop and $ T_{\mathrm{s}} $ is the sampling interval. Here, $ \myRound{\cdot} $ rounds to the closest integer. Based on the channel model given in (\ref{eq: channel model}) and the radar-transmitted signal given in (\ref{eq: transmitted signal p h m}), the UE-received signal can be modeled as
\begin{align}\label{eq: yh(i)}
	&~ {y}_h(i) = \mathbf{h}^{\mathrm{H}} \mathbf{s}_{h}(i) + \xi_h(i), \\
	\mathrm{s.t.}~&~ \mathbf{s}_h(i) = [s_{h0}(i)F_{h0}(i),\cdots,s_{h(M-1)}(i)F_{h(M-1)}(i)]^{\mathrm{T}}\nonumber
\end{align}
where $ (\cdot)^{\mathrm{H}} $ denotes the conjugate transpose, and $ \xi_h(i) $ is an AWGN.
Note that the above signal model is obtained based on a perfect timing, which may be challenging to realize in practice, as will be discussed in Section \ref{subsec: information demodulation}. Also note that $ F_{hm}(i) $ is multiplied to the radar signal to indicate the information embedded in the radar waveform. In different signaling schemes, $ F_{hm}(i) $ can vary, as reviewed in the following section.

\section{Overview of Existing Signaling Strategies for FH-MIMO DFRC}

Several signaling strategies have been specifically designed for FH-MIMO DFRC. Overall, the existing strategies can be categorized into two groups, one based on conventional phase modulations \cite{DFRC_AmbiguityFunc2018Amin,FH_MIMO_Radar2019_RadarConf,DFRC_DPSK2019Amin,DFRC_cskCPM} and the other exploiting the diversity in frequency hopping \cite{DFRC_FHcodeSel2018}.
In this section, the existing signaling strategies are first reviewed with their modulation and demodulation processing illustrated. Then the strategies are compared from various aspects.

\subsection{Information Modulation}\label{subsec: information modulation}

\subsubsection{Phase Modulations}
 In \cite{DFRC_AmbiguityFunc2018Amin,FH_MIMO_Radar2019_RadarConf}, phase shift keying (PSK) is introduced to FH-MIMO DFRC by embedding PSK phases into radar signals. 
 More specifically, a constant phase $ F_{hm}(i) = e^{\mj{\varpi}_{hm}} $ is taken in (\ref{eq: yh(i)}), where $ {\varpi}_{hm}\in \Omega_{J}~(J\ge 1) $ and $ \Omega_{J}=\left\{  0,\cdots,\frac{2\pi(2^{J}-1)}{2^{J}} \right\} $ is the $ {J} $-bit PSK constellation.
 In \cite{DFRC_DPSK2019Amin}, the differential PSK (DPSK) is proposed to replace PSK, so as to reduce range sidelobes and out-of-band transmission. 
 To further improve the spectral shape of the modulated radar signal, the continuous phase modulation (CPM) is introduced in FH-MIMO DFRC \cite{DFRC_cskCPM}. 
 The underlying principle is that signals with smoother modulating phase correspond to lower out-of-band transmission. 

\subsubsection{Hopping Frequency-based Modulation}
In \cite{DFRC_FHcodeSel2018}, the combinations of hopping frequencies are used to convey information bits, referred to as frequency hopping code selection (FHCS). As illustrated earlier, only $ M $ out of $ K $ sub-bands are selected each hop as hopping frequencies. Thus, there are $ C_{K}^M $ combinations of hopping frequencies, and the combinations can be employed to convey up to $ \myFloor{\log_2C_{K}^M}  $ number of information bits per radar hop, where $ \myFloor{\cdot} $ rounds towards the negative infinity. 

\subsubsection{Comparing Two Modulation Methods and Discussions}
Both modulation methods will lead to changes to the original FH-MIMO radar system, but in a different manner. The phase modulations change the phases of radar signals, and the phases can severely vary over antennas and hops, depending on the information bits to be sent. In contrast, FHCS leaves the phases unaltered but changes the hopping frequencies over hops.
A windfall of the phase modulations is that the periodic spikes in the sidelobe regions of the range ambiguity function are suppressed, which will be explained shortly.
This beneficial feature is not shared by FHCS-modulated waveform which has similar range ambiguity function to the original radar waveform, since both waveforms are essentially based on randomly selected hopping frequencies. Nevertheless, a prominent advantage of FHCS over phase modulations is the reduced complexity in information demodulation and the improved robustness against channel estimation errors, as will be illustrated in Section \ref{subsec: information demodulation}.

\begin{figure}[!t]
	\centering
	\includegraphics[width=80mm]{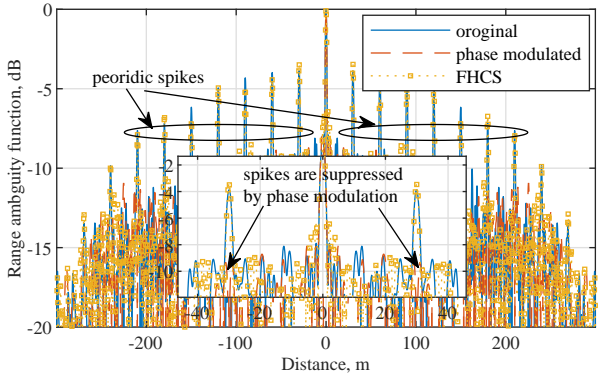}
	\caption{Comparing the range ambiguity functions of an FH-MIMO radar under different information modulations, where BPSK is used for the phase modulated waveform and the simulation parameters are set with reference to \cite{DFRC_FHcodeSel2018}. Specifically, $ H=M=10 $, $ B=100 $ MHz, $ K=20 $, $ T=0.2\mu $s and $ T_{\mathrm{s}}=1/(10B) $. A large sampling frequency is taken so that the curves look smoother.}
	\label{fig: range amb func comparisons}
\end{figure}

 Fig. \ref{fig: range amb func comparisons} compares the range ambiguity functions of an FH-MIMO radar under different modulations. We see the aforementioned spikes in both the original radar waveform and the FHCS-modulated one. These spikes are attributed to the re-use of sub-bands as hopping frequencies over hops; refer to \cite{DFRC_AmbiguityFunc2018Amin} for an in-depth analysis. 
We also see from Fig. \ref{fig: range amb func comparisons} that the sidelobe spikes are suppressed by the BPSK modulation. 
This is because BPSK scrambles the phases of the original radar waveform over hops and antennas, hence preventing the coherent accumulation in the sidelobe regions of the range ambiguity function.

\subsection{Information Demodulation}\label{subsec: information demodulation}

We proceed to describe the methods that demodulate the information embedded in radar signals. As seen from (\ref{eq: yh(i)}), the signals transmitted by different antennas of the radar transmitter are superimposed in the time domain. Nevertheless, based on the waveform orthogonality as depicted in (\ref{eq: waveform orthogonlity}), the signals can be separated in the frequency domain. 
The $ L $-dimensional discrete Fourier transform (DFT) of $ y_h(i) $ yields
\begin{align} \label{eq: Yh(l)}
	Y_{h}(l) = \beta_0 e^{-\mj m u_0} e^{\mj{\varpi}_{hm}} \delta(l-l_{hm}^*),~l_{hm}^*=k_{hm}\Delta,
\end{align}
where $ P=1 $ is taken as in prior works~\cite{DFRC_AmbiguityFunc2018Amin,FH_MIMO_Radar2019_RadarConf,DFRC_DPSK2019Amin,DFRC_cskCPM,DFRC_FHcodeSel2018,DFRC_pskDPSKcpm2020,Kai_channelDFRC_ICC2020} and $ \delta(l) $ is the Dirac delta function. The noise term is dropped for illustration convenience.

\subsubsection{FHCS Demodulation} \label{subsubsec: FHCS demodulation}
Based on $ Y_{h}(l) $, an FHCS symbol can be readily demodulated by identifying the peaks in the magnitude of $ Y_{h}(l) $. The indexes of the peaks result in the estimated set of $ \{{l}_{hm}^*~\forall m\} $. Based on the relation between $ l_{hm}^* $ and $ k_{hm} $ given in (\ref{eq: Yh(l)}), we can obtain the combination of hopping frequencies used in the radar hop. The estimated combination is then used to demodulate the FHCS symbol.

\subsubsection{PSK Demodulation}
Given the simple signal model,
PSK is used to illustrate the demodulation of the phase modulation-based signaling schemes; while DPSK and CPM can be demodulated similarly. Provided $ k_{hm} $ is known by the UE, $ l_{hm}^* $ is then obtained based on (\ref{eq: Yh(l)}). Extract the value of $ Y_h(l) $ at $ l=l_{hm}^* $, leading to 
\begin{align}\label{eq: Yhm tilde }
	\tilde{Y}_{hm} = Y_h(l_{hm}^*) = \beta_0  e^{-\mj m u_0} e^{\mj{\varpi}_{hm}}.
\end{align}
Provided that channel parameters are known, we can 
suppress $ \beta_0  e^{-\mj m u_0} $ and demodulate the PSK symbol $ {\varpi}_{hm} $. 

\subsubsection{Discussions}
\label{subsubsec: discussions in demodulation}

From the above description, the following information ($\mathcal{I}1$)-($\mathcal{I}3$) are necessary for demodulating information symbols in FH-MIMO DFRC: 
\begin{enumerate}
	\item[($\mathcal{I}1$)] Accurate timing;
	
	\item[($\mathcal{I}2$)] $ k_{hm} $ --- the hopping frequency used by antenna $ m $ at hop $ h $; and
	
	\item[($\mathcal{I}3$)] $ \beta_0 e^{-\mj m u_0} $ --- the channel response.
\end{enumerate}
Note that ($ \mathcal{I}1 $) is applied to obtain (\ref{eq: Yh(l)}) from (\ref{eq: yh(i)}) so that inter-hop and inter-antenna interferences are avoided. To this end, both FHCS and PSK demodulations implicitly apply ($\mathcal{I}1$), since the demodulations are performed based on (\ref{eq: Yh(l)}). 
As for ($\mathcal{I}2$) and ($\mathcal{I}3$), they are not required for FHCS demodulation and are explicitly used in the PSK demodulation.

Note that none of the information ($\mathcal{I}1$)-($\mathcal{I}3$) is easy to acquire in practice. For the first two information ($\mathcal{I}1$) and ($\mathcal{I}2$), a specific synchronization link may be required to establish between the radar and UE. Such a link can make the radar require extra resources, e.g., hardware and frequency etc. More specifically, an extra antenna and correspondingly RF chain can be required for synchronization, since the existing antennas of the radar transmitter are fully occupied for radar detection. Besides, a different frequency band (from the original radar operating frequency) can be required to establish the synchronization link so as to reduce the mutual interference between radar and communications. 
The acquisition of the third information ($\mathcal{I}3$) is subject to that of the first two, since inter-antenna and inter-hop interference will be incurred without an accurate timing.

\begin{table*}[!t]
	\centering
	\caption{Comparing Existing Signaling Schemes for FH-MIMO DFRC.}
	\label{tab: comparison of different signling schemes}
	{
		\begin{tabular}{c|c|c|c|c}
			\hline
			Modulation	&  PSK \cite{FH_MIMO_Radar2019_RadarConf} & DPSK \cite{DFRC_DPSK2019Amin} & CPM \cite{DFRC_cskCPM} & FHCS \cite{DFRC_FHcodeSel2018}\\
			\hline 
			Data Rate & PRF$\times HM J $ & PRF$\times HM J $ & PRF$\times H M\log_2M $ & PRF$\times H \myFloor{\log_2C_{K}^M} $ \\
			\hline		
			Range Sidelobe & low & lower & lowest & high with periodic spikes \\
			\hline
			Spectral Containment & good & better & best & poor \\
			\hline
			\makecell[c]{Necessary Information for\\Demodulation}  & ($\mathcal{I}1$), ($\mathcal{I}2$) and ($\mathcal{I}3$) & ($\mathcal{I}1$) and ($\mathcal{I}2$) & ($\mathcal{I}1$), ($\mathcal{I}2$) and ($\mathcal{I}3$) & ($\mathcal{I}1$) (not strictly necessary) \\
			\hline
		\end{tabular}
	}		
\end{table*}

\subsection{Comparing Signaling Schemes for FH-MIMO DFRC}\label{subsec: comparing signaling schemes}

The aforementioned signaling schemes are compared in terms of their impact on radar performance, their communication performance and the necessary information for demodulation.
The comparison is summarized in Table \ref{tab: comparison of different signling schemes}. 
Note that the radar ambiguity functions and spectral containment of the signaling schemes have been comprehensively compared in \cite{DFRC_pskDPSKcpm2020}. The data rates of different schemes are provided in the works \cite{DFRC_AmbiguityFunc2018Amin,FH_MIMO_Radar2019_RadarConf,DFRC_DPSK2019Amin,DFRC_cskCPM,DFRC_FHcodeSel2018}. Yet, the necessary information is overlooked in most of the existing works which consider ideally known ($\mathcal{I}1$)-($\mathcal{I}3$).

We notice that DPSK demodulation does not require ($\mathcal{I}2$) in quasi-static channels, since the channel response is suppressed in the differential processing \cite{DFRC_DPSK2019Amin}.  
In contrast to phase modulation-based schemes, FHCS does not require ($\mathcal{I}2$) or ($\mathcal{I}3$), since it pursues to estimate ($\mathcal{I}2$) from the magnitude of the frequency-domain signal, as illustrated in Section \ref{subsubsec: FHCS demodulation}. 
Strictly speaking, FHCS can be demodulated without requiring ($\mathcal{I}1$), provided that the timing offset is small. However, inter-hop and inter-antenna interferences are introduced given a non-zero timing offset, which consequently degrades the demodulation performance of FHCS. 
As will be shown in Section \ref{subsec: performance illustration}, the demodulation performance of FHCS has a significant impact on that of PSK. To this end, accurate timing is still desired for FHCS to attain a high communication performance of FH-MIMO DFRC.

We also notice that, unlike the phase modulation schemes having monotonically increasing data rates with respect to (w.r.t.) $ M $, the data rate of FHCS first increases with $ M $, then saturates as $ M $ reaches $ K/2 $, and starts decreasing as $ M $ keeps increasing, due to the binomial coefficient in its rate expression. {To improve the data rate, FHCS can be employed in combination with the phase modulation schemes, as developed in \cite{Kai_channelDFRC_ICC2020,Kai_JCASchannelTCOM_ArXiv}. However, the cost of the rate improvement is that more necessary information will be required for demodulation. This leads to the trade-off between data rate and demodulation performance in FH-MIMO DFRC.}

As for the demodulation performance of different singling schemes developed for FH-MIMO DFRC, no analytical result has been published yet.
In conventional communications, the asymptotic SER of DPSK and CPM is generally inferior to that of PSK, particularly when the modulation order is high; and the optimal receiver of CPM can be more difficult to achieve compared with the other two phase modulations \cite{book_simon2005commPerfmanceFading}. 
These general results, however, may not hold for FH-MIMO DFRC, since the phase demodulation performance is subject to the quality of ($\mathcal{I}2$) acquired in practice. 
As will be validated in Section \ref{subsec: performance illustration}, this coupling will significantly affect the demodulation performance of PSK.

\section{Channel Estimation for FH-MIMO DFRC}
\label{sec: channel estimation}

As illustrated above, most existing signaling schemes for FH-MIMO DFRC require accurate channel information to perform demodulation. Yet, to the best of our knowledge, only a few published work \cite{Kai_channelDFRC_ICC2020,Kai_JCASchannelTCOM_ArXiv} develop channel estimation scheme for FH-MIMO DFRC. In this section, we first review the channel estimation methods designed for DFRC (not based on FH-MIMO radar), and then briefly introduce the solution developed in \cite{Kai_channelDFRC_ICC2020}.

\subsection{Overview of Channel Estimations in DFRC}
To the best of our knowledge, few published works have explicitly dealt with the channel estimation issue of DFRC. Specifically, sparse recovery-based channel estimation methods are developed in  \cite{DFRC_ChnnlEst_sparse2016,DFRC_ChnnlEst_sparse2017} which coordinate radar and communication receiver using probing beams.
In a different yet relevant context (spectrum sharing), interference channel between radar and communication is estimated to achieve co-existence; refer to \cite{FanLiu2019Overview} for a review on those methods.  
In some recent DFRC works \cite{FanLiu2019Overview,Kai_rahman2020enablingSurvey}
channel estimation methods are developed, which, however, are based on new (future) DFRC waveforms/platforms.
These new designs are specifically tailored for dual functions and are non-trivial to be applied in FH-MIMO DFRC.
A common feature captured by most of the above methods is the full cooperation between radar and communication. 
In contrast, the channel estimation for FH-MIMO DFRC is likely to be carried out at a low-profile communication receiver with low computing power and a small number of (or a single) antennas. 

\subsection{Recent Advances in Channel Estimation for FH-MIMO DFRC} \label{subsec: channel estimation method ICC}
In \cite{Kai_channelDFRC_ICC2020}, a channel estimation scheme is developed for FH-MIMO DFRC to estimate the hopping frequencies and channel parameters, assuming an accurate timing. 

\subsubsection{Estimating Hopping Frequency}
\label{subsubsec: estimating hopping frequencies}
A novel FH-MIMO waveform with minor revisions to conventional ones is designed in \cite{Kai_channelDFRC_ICC2020}, which enables UE to estimate $ k_{hm} $, i.e., the hopping frequency used by the $ m $-th radar transmitter at any hop $ h $. 
In particular, it is proposed that:``\textit{the radar first randomly selects available sub-bands as hopping frequencies, as it usually does, then re-orders the frequencies in a deterministic order, ascending for instance, before assigning to each transmitter antenna.}''
The waveform re-ordering is illustrated in Fig. \ref{fig: signal model}. This simple processing enables $ k_{hm} $ to be estimated at the UE (instead of being acquired from the radar).

\begin{figure}[!t]
	\centerline{\includegraphics[width=80mm]{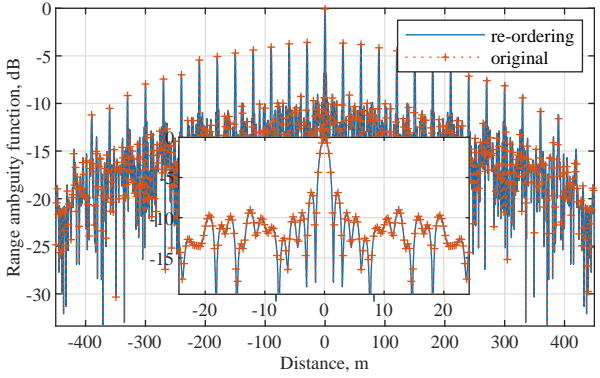}}
	\caption{Impact of the waveform re-ordering on the range ambiguity function. The same parameters as set for Fig. \ref{fig: range amb func comparisons} are used here.}
	\label{fig: range ambiguity function novel waveform}
\end{figure}

According to the novel waveform illustrated above, $ k_{hm}<k_{hm'}~(\forall m,m',m<m' )$ is ensured. 
This relation, applied in the set of sub-band indexes obtained in Section \ref{subsubsec: FHCS demodulation}, leads to an estimate of $ k_{hm}~\forall m $. 
Importantly, it is also proved in \cite{Kai_channelDFRC_ICC2020} that the above re-ordering does not incur any change to the range ambiguity function of the FH-MIMO radar. Fig. \ref{fig: range ambiguity function novel waveform} illustrates the range ambiguity function with or without performing the re-ordering. Clearly, the two range ambiguity functions overlap with each other.

\subsubsection{Estimating Channel Response}\label{subsubsec: channel estimation method}

To perform channel estimation, the first hop at each pulse is designated for channel estimation~\cite{Kai_channelDFRC_ICC2020} and hence has no information embedding, i.e., $ \varpi_{0m}=0~\forall m $. This, 
substituted in (\ref{eq: Yhm tilde }), leads to $ \tilde{Y}_{0m} = \beta_0  e^{-\mj m u_0} $. 
The path AoA $ u_0 $ can be regarded as a discrete frequency of a single-tone signal. 
By taking the DFT of $ \tilde{Y}_{0m} $ (over $ m $), the integer part of $ u_0 $ can be identified as the index of the strongest peak of $ |Z_{m'}| $, where $ Z_{m'} $ is calculated as
\begin{align}\label{eq: zm'}
	{Z}_{m'} = \sum_{m=0}^{{M}-1}\tilde{Y}_{0m} e^{-j\frac{2\pi m m'}{{M}}},~m'=0,1,\cdots,{M}-1.
\end{align}
Let $ \tilde{m} $ denote the index of the peak. A coarse estimation of $ u_0 $ is obtained as $ {2\pi \tilde{m}}/{M} $ whose estimation error is up to half of the frequency bin, i.e., $ {\pi}/{{M}} $.
To improve the estimation, interpolating DFT coefficients has been suggested in the literature \cite{FreqEst2019TCOM_Serbes,Kai_freqEst2020CL}. Next, we summarize the steps for $ u_0 $ estimation based on the latest $ q $-shift estimator (QSE), originally designed in \cite{FreqEst2019TCOM_Serbes} and improved in \cite{Kai_freqEst2020CL}.  
The technical details are suppressed here (interested readers are referred to \cite{FreqEst2019TCOM_Serbes,Kai_freqEst2020CL}).

To refine the estimate of $ u_0 $, QSE interpolates the DFT coefficients around $ \tilde{m} $, yielding
\begin{align}\label{eq: z_pm}
	Z_{\pm} = \sum_{m=0}^{{M}-1} \tilde{Y}_{0m} e^{-j\frac{2\pi m(\tilde{m}+\delta\pm\epsilon)}{{M}}},
\end{align}
where $ \delta $ is an interpolation factor to be updated iteratively and $ \epsilon $ is an intermediate variable satisfying $ \epsilon\le \min\{{M}^{-\frac{1}{3}},0.32\} $ \cite[Eq. (23)]{Kai_freqEst2020CL}. Constructing a ratio based on $ Z_{\pm} $, as given by $ \gamma=\frac{Z_{+}-Z_{-}}{Z_{+}+Z_{-}} $ and using the ratio $ \gamma $ to update $ \delta $ lead to
\begin{align}
	\delta = \frac{\epsilon\cos^2(\pi\epsilon)}{1-\pi\epsilon\cot(\pi\epsilon)}\times \Re\{\gamma\}+\delta,
\end{align}
where $ \delta $ on the right-hand side (RHS) of the equality is the old value, and $ \Re\{\cdot\} $ takes the real part of a complex number. By updating $ \delta $ for three times, 
the algorithm can generally converge \cite{FreqEst2019TCOM_Serbes}. 
The final estimate of $ u_0 $ and $ \beta_0 $ can be given by,
\begin{align}\label{eq: beta estimate}
	\hat{u}_0 = \frac{2\pi}{M}(\tilde{m}+\delta);~\hat{\beta}_0 = \frac{1}{{M}}\sum_{m=0}^{{M}-1} \tilde{Y}_{0m} e^{j m\hat{u}_0 }.
\end{align}

\subsection{Performance Illustration}\label{subsec: performance illustration}
\begin{figure}[!t]
	\centerline{\includegraphics[width=80mm]{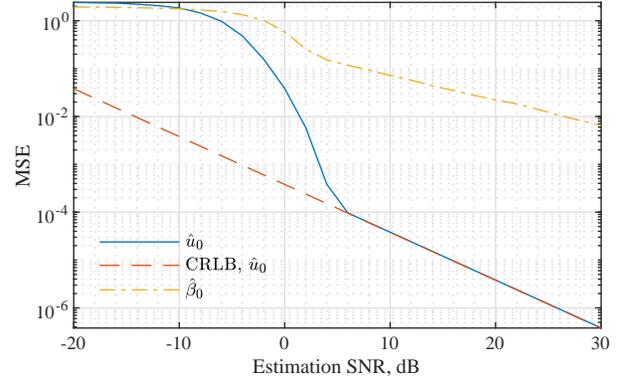}}
	\caption{Illustration of the channel estimation performance using the method elaborated on in Section \ref{subsec: channel estimation method ICC}. The same parameters as set for Fig. \ref{fig: range amb func comparisons} are used here, except that the sampling frequency is changed to twice the bandwidth. In addition, the AoD of the LoS path is set as $ u_0=\pi/2$; the path gain $ \beta_0 $ has unit amplitude but random angles uniformly taken in $ [0,2\pi] $ radius.}
	\label{fig: mse AoA vs snr}
\end{figure}

Numerical results are provided here to demonstrate the high estimation performance of the method elaborated on above. 
Fig. \ref{fig: mse AoA vs snr} plots the mean squared error (MSE) of the estimated channel parameters against the SNR, as measured based on (\ref{eq: yh(i)}) and denoted by $ \gamma_0 $ below. 
To evaluate the MSE of $ \hat{u}_0 $, $ N=2\times 10^4 $ independent trials are performed, and the MSE is calculated by $ \sum_{n=0}^{N-1}(\hat{u}_0(n)-u_0)^2/N $, where $ \hat{u}_0(n) $ is an estimate of $ u_0 $
and is obtained in the $ n $-th independent trial. The MSE of $ \hat{\beta}_0 $ is calculated similarly. Based on \cite{FreqEst2019TCOM_Serbes}, the Cram\'er-Rao lower bound (CRLB) of $ u_0 $ estimation is given by $ 3/{(\pi^2 ML \gamma_0 )} $.

We see from Fig. \ref{fig: mse AoA vs snr} that the MSEs of both $ \hat{u}_0 $ and $ \hat{\beta}_0 $ monotonically increase with the SNR. In particular, the MSE of $ \hat{u}_0 $ approaches the CRLB when the SNR exceeds $ 6 $ dB, which validates the high accuracy of the estimation method provided in Section \ref{subsubsec: channel estimation method}. A noticeable gap exists between the MSE of $ \hat{u}_0 $ and the CRLB. This is caused by the threshold effect when detecting the DFT peak for a coarse $ u_0 $ estimate. Please refer to \cite{FreqEst2019TCOM_Serbes} for more details on this effect.

\begin{figure}[!t]
	\centering
	\includegraphics[width=80mm]{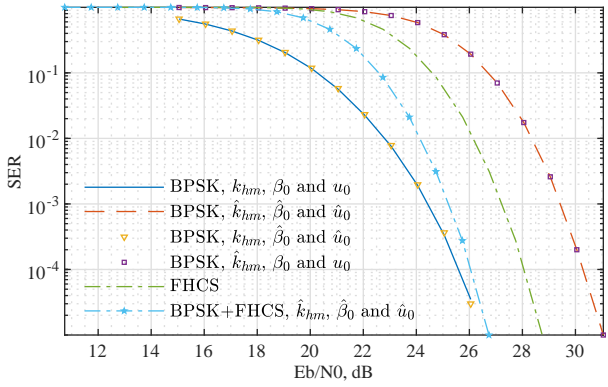}
	\caption{Illustration of the SER performance of FH-MIMO DFRC under different modulations, where $ \hat{k}_{hm} $ is the estimated hopping frequency used by antenna $ m $ at hop $ h $; $ \hat{\beta}_0 $ and $ \hat{u}_0 $ are obtained at the estimation SNR of $ 10 $ dB; and $ k_{hm} $, $ \beta_0 $ and $ u_0 $ are the true values. Other parameters are set the same as those used in Fig. \ref{fig: mse AoA vs snr}.}
	\label{fig: ser2snr}
\end{figure}

Fig. \ref{fig: ser2snr} plots the SER of BPSK and FHCS-based FH-MIMO DFRC, where the channel estimates obtained at the estimation SNR of $ 10 $ dB are used to perform information demodulations.
The $ x $-axis of the figure is based on the energy per bit to noise
power density ratio, which is calculated by $ LM\gamma_0BT/\tilde{J} $, where $ \tilde{J} $ is the number of bits conveyed per radar hop. Specifically, $ \tilde{J}=M=10 $ for BPSK $ \tilde{J}=\myFloor{\log_2C_{K}^M}=17 $ for FHCS, and $ \tilde{J}=10+17=17 $ for the combination of FHCS and BPSK. 

We see from Fig. \ref{fig: ser2snr} that, with either the estimated $ k_{hm} $ or its true value, the SER performance of BPSK achieved based on the estimated channel parameters approaches the performance under the ideally known channels. This again validates the high accuracy of the channel estimation method. 
We also see from the figure that, employing the estimated channel parameters, FHCS outperforms BPSK w.r.t the SER against $ \mathrm{E}_\mathrm{b}/\mathrm{N}_0 $, which owes to the improved data rate of FHCS over BPSK. Since the combination of FHCS and BPSK further improves the data rate, the SER performance of the combination is improved over those of the individual modulations.

We further see from Fig. \ref{fig: ser2snr} that the $ k_{hm} $ has a much stronger impact on the SER of BPSK compared with the channel parameters $ \beta_0 $ and $ u_0 $. To achieve the SER of $ 10^{-4} $, using the true value of $ k_{hm} $ can reduce the required $ \mathrm{E}_\mathrm{b}/\mathrm{N}_0 $ by $ 5 $ dB, compared with using the estimated $ k_{hm} $. On one hand, this observation indicates that the final SER performance of FH-MIMO DFRC should be carefully analyzed after taking the detection performance of $ k_{hm} $ into account, as discussed in Section \ref{subsec: comparing signaling schemes}. On the other hand, this observation strongly suggests that new methods should be developed to improve the estimation performance of $ k_{hm} $.

\section{Major Research Challenges in FH-MIMO DFRC}

As a newly conceived DFRC architecture, there are still many challenging issues to be addressed effectively. In this section, we discuss the major challenges and suggest potential solutions for future research.

\subsection{New Signaling Strategies Enabling Higher Data Rates} \label{subsec: new signling and higher data rate}
As shown in Section \ref{subsec: performance illustration}, the overall data rate achieved by FH-MIMO DFRC is $ 27 $ Mbps, under the condition that BPSK and FHCS is combined, the radar has $ 10 $ antennas and the bandwidth of $ 100 $ MHz, and the UE has a single antenna. 
To provide an effective high-speed G2A communication link using FH-MIMO DFRC, new signaling strategies are required.

We remark that the information conveyed in the hopping frequencies has not been fully explored yet.
In light of FHCS as illustrated in Section \ref{subsec: information modulation}, the permutations of hopping frequencies can be employed to convey information bits \cite{Kai_JCASsecureTWC_ArXiv}.
Using permutations is expected to increase the data rate dramatically, since the number of permutations can be huge. 
Take the parameters used in Fig. \ref{fig: ser2snr} for an example. 
The number of permutations that permutes $ M(=10) $ out of $ K(=20) $ sub-bands can be larger than $ 670 $ billion, which, in theory, is able to convey up to $ 39 $ bits per radar hop. As a result, an increase of $ 21 $ Mbps in data rate can be achieved through using permutations.  
This improvement of data rate, however, requires a proper demodulation method to be developed at the UE. One challenge is to deal with the reduced Euclidean distances between the permutations. {Another challenge is to develop effective receiving schemes for UE to reliably demodulate the permutation-based information symbol, particularly when the issues discussed sequentially are present.}

\subsection{Timing Offset}

\begin{figure}
	\centering
	\includegraphics[width=80mm]{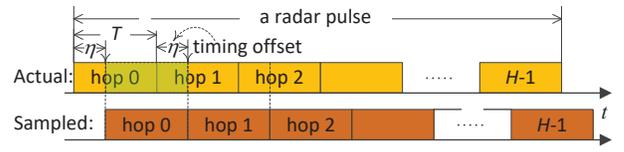}
	\caption{Illustration of the impact of a non-zero timing offset on the signal sampling at a UE.}
	\label{fig: timing offset}
\end{figure}

The prior works on FH-MIMO DFRC \cite{DFRC_AmbiguityFunc2018Amin,FH_MIMO_Radar2019_RadarConf,DFRC_DPSK2019Amin,DFRC_cskCPM,DFRC_FHcodeSel2018,DFRC_pskDPSKcpm2020,Kai_channelDFRC_ICC2020} assume perfect timing which is also the case in our elaboration provided in Section \ref{sec: channel estimation}. However, there can be a non-zero timing offset in practice, particularly when a synchronization link between the radar and the UE is unavailable; see Section \ref{subsec: information demodulation} for details.  
In a packet-based communication, the UE can attain a coarse timing by performing conventional methods like energy or correlation-based detection~\cite{Kai_JCASchannelTCOM_ArXiv}. Provided a non-zero timing offset, a sampled hop spans over two actual hops, as illustrated in Fig. \ref{fig: timing offset}. As a consequence, there will be inter-hop interference introduced to the received signals. Moreover, due to the incomplete sampling of a hop, the waveform orthogonality of the FH-MIMO radar, see (\ref{eq: waveform orthogonlity}), is destroyed, hence introducing inter-antenna interference~\cite{Kai_JCASchannelTCOM_ArXiv}. 

It is noteworthy that the impact of timing offset is frequency-varying, which can result in great difficulty for channel estimation and information demodulation. 
Based on the baseband radar signal given in (\ref{eq: transmitted signal p h m}), we can model the received signal with a non-zero timing offset, as given by $ e^{\mj2\pi (i-L_{\eta})k_{hm} \Delta} $, where $ L_{\eta} $ is the number of samples in the non-zero timing offset denoted by $ \eta $. 
Since $ k_{hm} $ varies across hop $ h $ and antenna $ m $, we notice that the phase nuisance incurred by $ L_{\eta} $ also varies over hops and antennas. To this end, the timing offset can lead to the following severe consequences:

\begin{enumerate}
	\item The antenna-varying phase disturbance will invalidate many angle estimation methods, including the one elaborated on in Section \ref{subsec: channel estimation method ICC}, that rely on the linear phase relation across the antennas of radar transmitter;
	
	\item The DFRC channel presents a fast fading from hop to hop, which dramatically burdens the channel estimation and communication overhead;
	
\end{enumerate}

A seemingly straightforward solution to this issue is to perform additional timing offset estimation, which helps recover a complete hop and revive the methods for channel estimation illustrated in Section \ref{sec: channel estimation}. 
However, the estimation of the timing offset is challenging.
As mentioned earlier, inter-antenna and inter-hop interference exists for any given non-zero timing offset. On the other hand, the antenna- and hop-varying phase resulted from the timing offset is coupled with the phase of channel response in a point-wise manner. 
This makes the estimation of either part challenging. To address the above challenges, we may resort to the degrees of freedom (DoF) available in the FH-MIMO radar waveform. 
For example, if $ k_{hm}=0 $, then the phase caused by the timing offset is removed according to $ e^{\mj2\pi (i-L_{\eta})k_{hm} \Delta} $. 
More works are demanded to address the issue of timing offset effectively.

\subsection{Multi-Path Channel}

\begin{figure*}[!t]
	\centerline{\includegraphics[width=120mm]{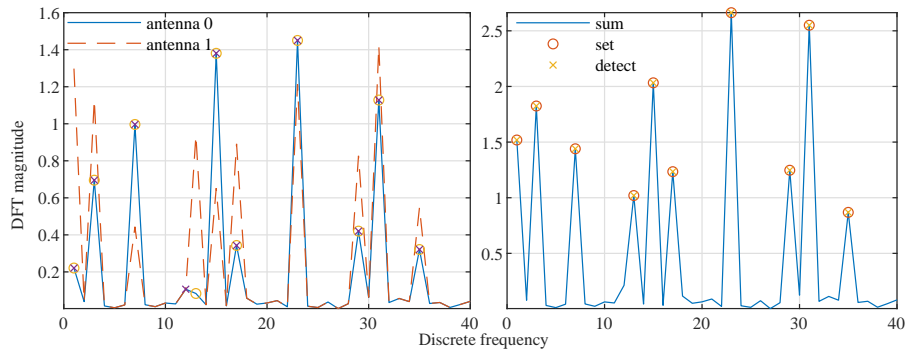}}%
	\caption{Illustration of the severe signal attenuation caused by multi-path fading, where the Rician channel is considered with the Rician factor of $ 5 $ dB and the LoS path takes the unit power. 
		Two antennas with the spacing of half a wavelength are considered at the UE, each with an independent AWGN of $ -10 $ dB noise power. 
		The radar parameters are set the same as those in Fig. \ref{fig: range amb func comparisons}. }
	\label{fig: signal cancellation}
	\vspace{-10pt}
\end{figure*}

Thus far, only the LoS-dominated AWGN channel is considered for FH-MIMO DFRC. As discussed in Section \ref{subsec: channel model}, it can be practical to consider the LoS channel for some scenarios, such as the G2A communications between a ground-based radar and an HAP; see Fig. \ref{fig: DFRC system}. However, as also illustrated in the figure, multi-path fading can happen in FH-MIMO DFRC, when the airborne UE, e.g., UAV, flies at a low altitude. 
Under the multi-path fading, the signal transmitted by some antennas of the radar transmitter can be severely attenuated.

Fig. \ref{fig: signal cancellation} illustrates the signal attenuation caused by multi-path fading by plotting $ |Y_h(l)| $, namely, the magnitude of the DFT of the UE-received signal within a radar hop, where $ Y_h(l) $ is given in (\ref{eq: Yh(l)}) and an accurate timing is assumed here. Based on the waveform design illustrated in Section \ref{subsubsec: estimating hopping frequencies}, the $ m $-th DFT peak corresponds to the signal from antenna $ m $~$ (m=0,1,\cdots,M-1) $.  
We see from the figure that a severe attenuation of the signal transmitted by antenna three is observed on the first UE antenna. Performing the method elaborated in Section \ref{subsubsec: estimating hopping frequencies} to detect the hopping frequency, we see from the left subfigure in Fig. \ref{fig: signal cancellation} that the hopping frequency of the third antenna is incorrectly detected. 
This incorrect detection will lead to incorrect demodulation of not only hopping frequency-based modulations, e.g., FHCS, but also phase-based modulations, e.g., PSK. For the later modulation, the signal used for demodulation is extracted based on the identified indexes of the DFT peaks.

A viable solution to combating the multi-path fading in FH-MIMO DFRC is equipping the UE with multiple antennas and hence exploit the antenna diversity. 
This is demonstrated in Fig. \ref{fig: signal cancellation}. The left subfigure also plots the magnitude of the DFT of the signal received by another UE antenna. Clearly, the multi-fading attenuation of the signal transmitted by the third antenna is greatly relieved. By combining the signals of the two antennas incoherently, the sum signal can be used to detect the hopping frequency more reliably; see the right subfigure in Fig. \ref{fig: signal cancellation}. 
It is worth mentioning that improving the detecting performance of the hopping frequencies also substantially benefits to the SER performance of demodulating phase-based modulations, as demonstrated in Fig. \ref{fig: ser2snr}.  
With the benefits of using multi-antenna receivers at the UE recognized, we should also bear in mind that effective signal reception also relies on the accurate estimation of timing offset and channel parameters. These practical issues have not been properly addressed for a single-antenna receiver yet, not to mention multi-antenna configurations. 

\subsection{Security Issues}
FH-MIMO DFRC can have a serious security issue, since the underlying FH-MIMO radar has an omnidirectional radiation pattern in power \cite{Kai_JCASsecureTWC_ArXiv}.
Provided that the signal framework used between the radar and the legitimate UE is known by an eavesdropper, the radar-transmitted signal can be readily demodulated by an eavesdropper using the method provided in Section \ref{sec: channel estimation}. 
As an effective supplement to
upper-layer security techniques \cite{Kai_PLSlensArray}, physical layer security has attracted increasing attention recently in 5G and beyond communications, and can be a promising solution to achieving a secure FH-MIMO DFRC.

Specifically, the secrecy enhancement method developed in \cite{PLS_multiAntenna_robertHeath2017TVT} may be tailored for FH-MIMO DFRC. In short, the method randomly selects a subset of antennas in a transmitter array and half of the antennas in the subset have the signs of their beamforming weights reversed. Through the above processing, AWGNs are injected to the whole spatial region except for the direction of a targeted receiver. The method may be tailored for FH-MIMO DFRC, however, at the cost of reducing the data rate. This is a typical trade off between the secrecy rate and data rate of a communication system. 
To this end, the security issue can be treated in combination with designing new signaling strategies for high data rates, as discussed in Section \ref{subsec: new signling and higher data rate}.

\section{Conclusions}
In this article, a comprehensive review on FH-MIMO DFRC is provided. The potential applications and channel scenarios are presented. The existing signaling strategies 
for FH-MIMO DFRC are surveyed with their performance compared from various aspects. A channel estimation scheme specifically designed for FH-MIMO DFRC is presented with numerical results to demonstrate estimation and communication performance. Major issues remained in FH-MIMO DFRC are discussed with potential solutions suggested. We envision that, with the highlighted challenges properly addressed in the future, FH-MIMO DFRC is poised to become a viable solution to long-range radar and communications, and forthcoming ISTN and 6G mobile networks.

\bibliographystyle{IEEEtran}
\bibliography{IEEEabrv,../ref/bib_JCAS.bib}
\end{document}